\documentclass[twocolumn, pra, groupedaddress, superscriptaddress, showkeys, showpacs, notitlepage]{revtex4-1}

\usepackage{amsmath,amssymb}  
\usepackage{amsfonts}
\usepackage{graphicx}   
\usepackage{hyperref}   
\usepackage{color}

\newcommand*\pFq[2]{{}_{#1}F_{#2}}

\begin{document}

\title{
Soft topological modes protected by symmetry in rigid mechanical metamaterials}
\author{Hridesh Kedia}
\affiliation{School of Physics, Georgia Institute of Technology, Atlanta, GA 30332}
\affiliation{Physics of Living Systems Group, Department of Physics, Massachusetts Institute of Technology, Cambridge, MA 02139}

\author{Anton Souslov}
\affiliation{Department of Physics, University of Bath, Claverton Down, Bath BA2 7AY, UK}

\author{D.~Zeb Rocklin}
\email{zebrocklin@gatech.edu}
\affiliation{School of Physics, Georgia Institute of Technology, Atlanta, GA 30332}

\begin{abstract}
Topological mechanics can realize soft modes in mechanical metamaterials in which the number of degrees of freedom for particle motion is finely balanced by the constraints provided by interparticle interactions. However, solid objects are generally hyperstatic (or overconstrained).
Here, we show how symmetries may be applied to generate topological soft modes even in overconstrained, rigid systems. To do so, we consider non-Hermitian topology based on non-square matrices, and design a hyperstatic material in which low-energy modes protected by topology and symmetry appear at interfaces. Our approach presents a novel way of generating softness in robust scale-free architectures suitable for miniaturization to the nanoscale.
\end{abstract}
\maketitle

Topologically protected modes possess novel properties and extraordinary robustness stemming from their dual nature: these modes appear at the \emph{boundaries}, yet are generated by \emph{bulk} properties~\cite{huber2016topological,bertoldi2017flexible,mao2018maxwell,lu_topological_2014,ozawa2019topological,ashida2020,shankar2020,hasan2010colloquium}. 
First realized in electronic states~\cite{von_klitzing_quantized_1986,thouless_quantized_1982,haldane_model_1988,kosterlitz_ordering_1973,hasan2010colloquium}, this topological bulk-boundary correspondence has since been extended to the mechanics, acoustics, and photonics of structured matter~\cite{rechtsman_photonic_2013,lu_topological_2014,ozawa2019topological,nash_topological_2015,wang_topological_2015,peano_topological_2015,susstrunk_observation_2015,khanikaev_topologically_2015,paulose2015selective,he_acoustic_2016,chen_tunable_2016,chen2016topological,souslov_topological_2017,murugan_topologically_2017,dasbiswas_topological_2018,vila_observation_2017,trainiti_optical_2017,mitchell_amorphous_2018,souslov2019topological,susstrunk_classification_2016,huber2016topological,bertoldi2017flexible,mao2018maxwell,Tauber2020}. 
All of these systems are characterized by \emph{topological invariants}, quantized numbers associated with a physical state. 
The type of invariants that a particular system can exhibit, or whether such topological character can exist at all, depends on its symmetries and has been classified for conventional, Hermitian Hamiltonians via the \emph{tenfold way}~\cite{ryu_topological_2010}. More recently, topological concepts have been generalized to open quantum and classical systems in the presence of external drive and dissipation using fundamental ideas from non-Hermitian physics, which nevertheless focus on square Hamiltonians~\cite{yao_edge_2018,esaki_edge_2011,liang_topological_2013,lee_anomalous_2016,leykam_edge_2017,menke_topological_2017,xu_weyl_2017,gonzalez_topological_2017,xiong_why_2018,shen_topological_2018,hu_exceptional_2017,kawabata2019symmetry, yoshida_exceptional_2019,brandenbourger2019non,Baardink2020,scheibner2020,ashida2020}.

For the mechanics of ball-and-spring networks, Ref.~\cite{kane_topological_2014} predicts a number of localized floppy (zero-energy) modes proportional to a local flux of a bulk topological polarization. These topologically protected modes have been realized in mechanical metamaterials along interfaces in one, two, and three dimensions, as well as at dislocation defects~\cite{kane_topological_2014, chen2014nonlinear, rocklin2016mechanical, paulose2015topological, bilal2017intrinsically, baardink2018localizing}.
These realizations require a fine balance (called isostaticity) between the numbers of degrees of freedom and constraints (e.g., springs) to define the underlying topological polarization.
Isostaticity allows for this topological invariant by enforcing a one-to-one mapping between degrees of freedom at sites and the bonds between them. These distinct quantities can then be related via a non-Hermitian rigidity matrix (analogous to a Hamiltonian), which is square only for isostatic systems. However, isostatic materials are inherently unstable~\cite{guest2003determinacy} making them susceptible to deformations due to thermal fluctuations. This makes realizing isostatic topological lattices in atomic, molecular, or colloidal crystals especially challenging.

For the non-isostatic case, topologically protected soft modes are a consequence of topological invariants distinct from topological polarization. For example, overconstrained (i.e., hyperstatic) systems can possess low-dimensional topological boundary modes at the corners of two-dimensional systems~\cite{saremi_controlling_2018}. 
For modes at interfaces (instead of corners), Ref.~\cite{roychowdhury_classification_2018} includes an exhaustive classification scheme evocative of the tenfold way but for non-square non-Hermitian Hamiltonians, which could be applied to design non-isostatic systems. However, the lack of a bulk-boundary correspondence principle for the topological classification in Ref.~\cite{roychowdhury_classification_2018} leaves open the problem of realizing topologically protected interface modes in overconstrained systems.

In this letter, we focus on mechanically stable lattices, which are overconstrained, as are nearly all naturally occurring crystals. They are robust to thermal fluctuations and can be realized on atomic, molecular or colloidal scales. Generically, these crystals are not expected to have any soft modes. Building on the classification scheme in Ref.~\cite{roychowdhury_classification_2018}, we design materials in which a certain symmetry class can guarantee the presence of soft modes at any interface between topologically distinct states. We dub this the \emph{generalized inversion symmetry} and design a one-dimensional hyperstatic lattice which respects this symmetry. We show for the first time, using exact solutions and numerical calculations, that topological modes localized at interfaces between topologically distinct lattice configurations arise in an overconstrained mechanical system. Furthermore, we show that these topologically protected interface gap modes have low energies rather than the zero energy modes of Ref.~\cite{kane_topological_2014}. We show that at interfaces for which the bulk band-gaps on the two sides of the interface are sufficiently different, these topological modes are absent. Our work contributes to the understanding of non-Hermitian topology by extending design principles for topological modes beyond square Hamiltonians to rectangular matrices.

\begin{figure}[!t]
\centering
\includegraphics[width=\columnwidth]{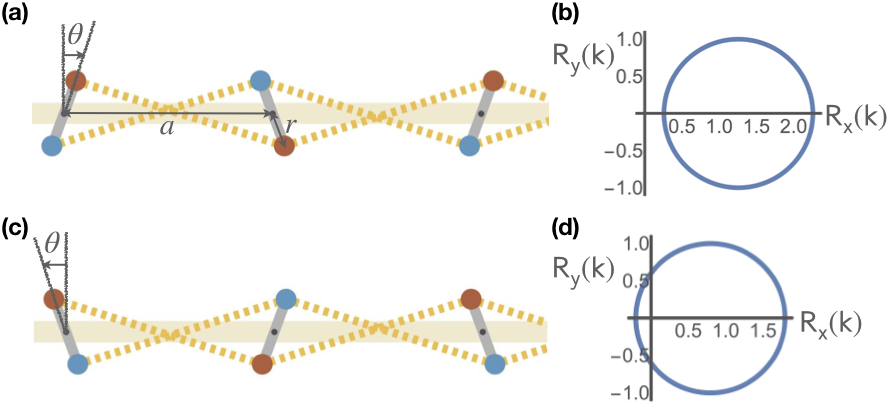}
\caption{(Color) Hyperstatic 1D rotor chain with generalized inversion symmetry. (a) Right-leaning rotor chain with positive rotor angle $\theta$ (measured from the red rotor head). (b) The winding number of $\vec{R}(k)$ around the origin is $0$ for $\theta > 0$. (c) Left-leaning rotor chain with negative $\theta$. (d) The winding number of $\vec{R}(k)$ around the origin is $1$ for $\theta < 0$.}
\label{hyperstatic_rotor_chain_intro}
\end{figure}

\textit{Generalized inversion symmetry.}---The linear deformations of a mechanical system may be described via a \emph{rigidity matrix} $\mathcal{R}$, a linear map $\mathbf{e} = \mathcal{R}\cdot \mathbf{u}$ determined from the system geometry that maps the displacements of sites $\mathbf{u}$ (or more general degrees of freedom) onto the extensions of springs $e$ (or more general violations of constraints) and hence may be used to generate a potential energy. In real space, this rigidity matrix is real but not necessarily square. {For periodic systems, the matrix can be written in Fourier space in terms of the wavenumber $k$, resulting in a block diagonal matrix with blocks $\mathcal{R}(k)$ as Laurent polynomials  in powers of the phase factor $\exp(i k)$ with real coefficients $\mathbf{R}_n$: $\mathcal{R}(k) = \sum_n \mathbf{R}_n \, \exp(ikn)$.} Zero edge modes appear at complex wavenumbers $k$, for which the phase factor $\exp(i k)$ becomes a general complex number $z$~\cite{lubensky2015phonons,mao2018maxwell}.

Generalized inversion symmetry is defined by the existence of a basis in which $\mathcal{R}(k)$ is real for real wavenumbers $k$, i.e., that there exist unitary matrices $U, W$ such that $U^\dagger\cdot\mathcal{R}(k)\cdot W$ is a real matrix for all real $k$. A consequence of this symmetry is that zero modes come in pairs: for every zero mode at complex number $z$, where $z=\mathrm{e}^{ik}$, there is also a zero mode at $z^{-1}$, as shown in the Appendix. In other words, a lattice with generalized inversion symmetry has equal numbers of zero modes localized on the left and the right interfaces. Lattices with generalized inversion symmetry can be classified into topologically distinct phases \cite{roychowdhury_classification_2018}, even when the number of constraints differs from the number of degrees of freedom.

The canonical example of a mechanical lattice with topologically protected modes: the one-dimensional chain of rotors and springs studied in Ref.~\cite{kane_topological_2014}, does not obey generalized inversion symmetry. 
Its zero mode is localized on either the right or the left interface indicating the presence of a topological polarization.

The topological invariant that distinguishes lattices in the classification that we use is calculated from the Singular Value Decomposition (SVD, a generalization of the eigenvalue decomposition) of the rigidity matrix. In SVD, the Fourier-transformed rigidity matrix $\mathcal{R}(k)$ is written as $\mathcal{R}=\mathcal{U}\Lambda_R\mathcal{V}^\dagger$, with $\mathcal{U},\mathcal{V}$ unitary and $\Lambda_R$ a rectangular matrix with only non-negative so-called \emph{singular values} along the diagonal. $\mathcal{R}(k)$ can be transformed into its SVD-flattened version $\mathcal{Q}(k)$ by replacing every nonzero element of $\Lambda_R$ by $1$. For an isostatic lattice, in the basis in which $\mathcal{R}(k)$ is real, its SVD-flattened version $\mathcal{Q}(k)$ is a real orthogonal matrix, which can be classified into topologically distinct classes according to the homotopy groups of such matrices. Even a hyperstatic lattice in which the number of constraints per unit cell exceeds the number of degrees of freedom per unit cell by one can be similarly classified. We do this by adding to the SVD-flattened rigidity matrix $\mathcal{Q}(k)$ a column that is orthogonal to all its column vectors and thus transforming $\mathcal{Q}(k)$ to be orthogonal.

\emph{Maxwell lattices.}--- So far, we have only considered the classification of rigidity matrices, in line with the matrix classification scheme from Ref.~\cite{roychowdhury_classification_2018}. Now, we proceed beyond classification to realising rigidity matrices for topological materials with generalized inversion symmetry.
For an \emph{isostatic} lattice with generalized inversion symmetry and $2$ sites per unit cell, the SVD-flattened rigidity matrix $\mathcal{Q}(k)$ is equivalent to a two-dimensional rotation matrix. The topological invariant classifying such a lattice is the integer winding number of its rotation angle around the unit circle, as the wavenumber $k$ goes from $0$ to $2\pi$. At an interface where this topological invariant changes, there appear topologically protected gap modes, as shown in the appendix. 

Similarly, for an isostatic lattice with generalized inversion symmetry and $N=3$ sites per unit cell, the SVD-flattened rigidity matrix $\mathcal{Q}(k)$ is equivalent to a three-dimensional rotation matrix. To calculate its topological invariant, we represent $\mathcal{Q}(k)$ by a point in a solid sphere of radius $\pi$ whose antipodal points are identified, where the radius vector of the point encodes the rotation angle in its magnitude and the rotation axis in its direction. Then the topological invariant is $0$ or $1$ depending on the contractibility of the loop traced out by $\mathcal{Q}(k)$, as the wavenumber $k$ goes from $0$ to $2\pi$. At an interface where this topological invariant changes, there appears a topologically protected gap mode, as described in the Appendix. Maxwell lattices obeying generalized inversion symmetry with more than $3$ sites per unit cell are similarly characterized by a $\mathbb{Z}_2$ topological invariant.

\begin{figure}[!tb]
\centering
\includegraphics[width=\columnwidth]{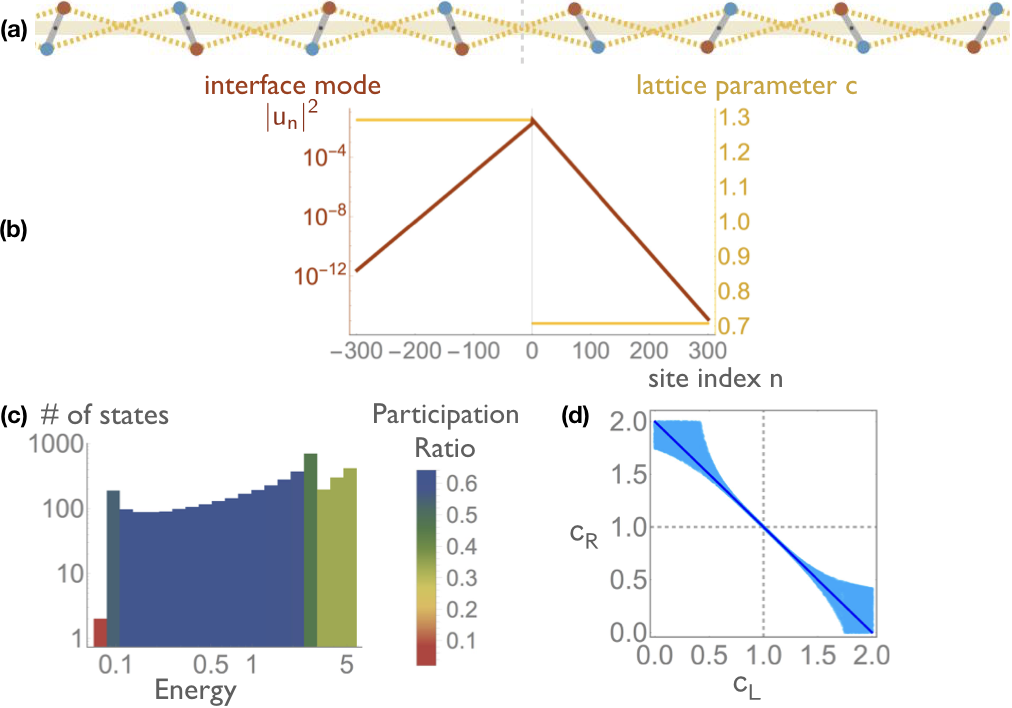}
\caption{(Color) Topologically protected localized modes at a sharp interface in a hyperstatic lattice with generalized inversion symmetry. (a) A sharp interface between right and left-leaning hyperstatic rotor chains. (b) The localized mode (red), the lattice parameter c (yellow) at the interface.  (c) Density of states showing two topologically protected localized modes, one at each interface. (d) The region of parameter space (light blue) for localized modes to exist at an interface between a lattice with $c=c_L$ on the left, and $c=c_R$ on the right, with $c_L+c_R=2$ in deep blue.}
\label{hyperstatic_rotor_chain_sharp_interface}
\end{figure}

\emph{Topological modes in a hyperstatic lattice.}---The study of topological modes has hitherto been almost exclusively in the realm of Maxwell lattices. As we proceed to show, the advantage of generalized inversion symmetry is that it allows us to construct \emph{hyperstatic} lattices that have topologically protected modes. We restrict our analysis to hyperstatic lattices with one degree of freedom and two constraints per unit cell. For these lattices, the topological invariant has a simple interpretation as the winding number around the origin of the real two-component vector $\vec{R}(k)$ as $k$ crosses the Brillouin zone, as shown in Fig.~\ref{hyperstatic_rotor_chain_intro}(b,d). For instance, consider a hyperstatic lattice with the following Fourier transformed rigidity matrix: $\mathcal{R}(k)=\left( c-\cos{k}\,,\, \sin{k} \right)^T$, where $c>0$ is a dimensionless parameter determined via the structure's geometry. Its winding number is $1$ or $0$ for $c<1$ and $c>1$ respectively. When $c=1$, the loop traced out by $\vec{R}(k)$ passes through the origin of the 2D plane, indicating that the lattice must have a bulk zero mode as it crosses over from one topological phase to the other.

The above rigidity matrix $\mathcal{R}(k)$ can be transformed to an equivalent rigidity matrix $\tilde{\mathcal{R}}(k)$ which is real in real space, as follows:
\begin{align}
\tilde{\mathcal{R}}(k) &= \frac{1}{\sqrt{2}}\begin{pmatrix} 1 & -i \\ 
1 & i \end{pmatrix}\cdot \mathcal{R}(k) = \frac{1}{\sqrt{2}}\begin{pmatrix} c-\mathrm{e}^{i\,k} \\ 
c-\mathrm{e}^{-i\,k} \end{pmatrix}. \label{real_space_real_rigidity_2x1_fourier}
\end{align}
The rigidity matrix $\tilde{\mathcal{R}}(k)$ is realized by the hyperstatic rotor chain in Fig.~\ref{hyperstatic_rotor_chain_intro}(a,c) where $c=(a+2r\sin\theta)/(a-2r\sin\theta)$, $a$ is the lattice spacing, $r$ is the distance between the fixed point and the rotor head, and $\theta$ is the rotor angle measured from the vertical. The hyperstatic rotor chain with $c>1$ ($ \Leftrightarrow \theta >0$) and $c<1$ ($\Leftrightarrow \theta <0$) belong to topologically distinct phases with winding numbers $0$ and $1$, respectively, as shown in Fig.~\ref{hyperstatic_rotor_chain_intro}. 

The equation of motion for the angular displacement $u_n$ of a rotor at lattice site $n$ becomes:
\begin{align}
\ddot{u}_n &= -(c_n-1)^2u_n +c_n ( u_{n+1} - 2 u_n +u_{n-1} ) + \nonumber \\
&\qquad \frac{c_{n+1}-c_n}{2} u_{n+1} -\frac{c_n - c_{n-1}}{2} u_{n-1}, \label{real_space_eom_R_2x1}
\end{align} 
where the dimensionless parameter $c$ is taken to be a function of the lattice site $n$ ($c \rightarrow c_n$). For a detailed derivation, see Appendix.

We now analytically and numerically study the modes localized at an interface between topologically distinct hyperstatic rotor chains. These interface modes lie in the gap, i.e., have energy lower than the minimum bulk mode energy $(c_n-1)^2$, but are not necessarily soft. The interface modes are guaranteed to be soft when the lowest bulk mode energy, $(c_n-1)^2$, is small. These interface modes are reminiscent of the zero mode that the hyperstatic rotor chain must pass through as it is continuously deformed from the topological phase on one side of the interface ($c>1$) to the topological phase on the other side of the interface ($c<1$).

\emph{Sharp interface.}---At a sharp interface between topologically distinct phases of the hyperstatic rotor chain, the parameter $c$ jumps from below $1$ to above $1$ or vice-versa. Using normal modes of the generalized Bloch form $u_n(t) = u_0\, z^n\,\mathrm{e}^{i\omega t}$, we exactly solve the equation of motion, Eq.~(\ref{real_space_eom_R_2x1}) (see Appendix for details). Using our exact solution, we calculate the values of $c$ on the left $(c_L)$ and right $(c_R)$ of the interface, for which the exact solution admits a localized interface mode. As shown in Fig.~\ref{hyperstatic_rotor_chain_sharp_interface}(d), we find that the existence of a topological interface mode in the gap is guaranteed only when the gap frequencies on both sides of the interface overlap. This overlap occurs when the energies for the lowest bulk modes on either side of the interface coincide: $(c_L -1)^2 = (c_R-1)^2$. For the interface to obey this condition and be topologically non-trivial (i.e. $c_L \neq c_R$), it must satisfy: $c_L + c_R = 2$. Taking $c_L=1+m_0\,,\,c_R=1-m_0$ and substituting in the exact solution, the energy of the localized mode is $\omega^2=m_0^2 - m_0^4/(4-m_0^2)$, which is lower than the lowest bulk mode energy: $(c-1)^2 = m_0^2$. Furthermore, the mode decay rates are: $z_L=(2+m_0)/(2+m_0-{m_0}^2)\,,\,z_R = (2-m_0-{m_0}^2)/(2-m_0)$. That is, for $\vert m_0 \vert <1$, the mode amplitude is right-growing on the left of the interface ($\vert z_L\vert >1$),  and right-decaying on the right side of the interface ($ \vert z_R\vert < 1$), and hence localized at the interface.
\begin{figure}[!t]
\centering
\includegraphics[width=\columnwidth]{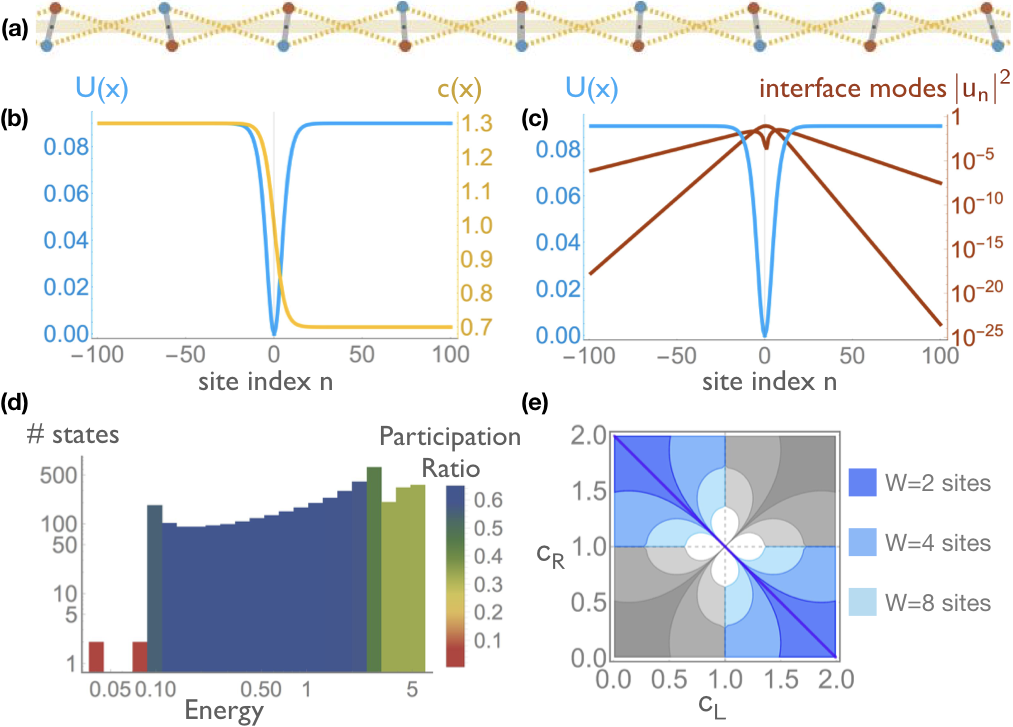}
\caption{(Color) Topologically protected localized modes at a smooth interface in a hyperstatic lattice with generalized inversion symmetry. (a) A smooth interface between right and left-leaning hyperstatic rotor chains. (b) For smoothly varying parameter $c$, the localized interface modes can be mapped to bound states in an effective potential (blue) for the Schr{\"o}dinger equation. (c) The two lowest localized modes (red) at the interface. (d) Density of states showing the two topologically protected modes at the $2$ interfaces. (e) The region of parameter space (blue) in which localized modes exist at a smooth interface with $c=c_L$ on the left, and $c=c_R$ on the right, calculated in the continuum approximation (see Ref.~\cite{kocher_criteria_1977} and Eq.~(C21) in the Appendix) for different interface widths $W$. $c_L+c_R=2$ is shown in deep blue. The grey regions correspond to the non-topological interface, when localized interface modes are not guaranteed to exist.}
\label{hyperstatic_rotor_chain_smooth_interface}
\end{figure}

To confirm our exact analysis, we numerically calculate the modes of a periodic hyperstatic chain of $N=4000$ rotors with two sharp interfaces. One of the interfaces with $c_L = 1.3\,,\, c_R = 0.7$ (i.e. $m_0 = 0.3$), is shown in Fig.~\ref{hyperstatic_rotor_chain_sharp_interface}(a). The topological mode localized at that interface is shown in Fig.~\ref{hyperstatic_rotor_chain_sharp_interface}(b), in red, with the mode decay rates in agreement with the calculated values: $z_L = 1.04$ and $z_R = 0.95$. The density of states shown in Fig.~\ref{hyperstatic_rotor_chain_sharp_interface}(c) shows the two topological modes localized at the two interfaces as having the lowest energy (in red), in agreement with the calculated value: $\omega^2 = 0.0879$. They also have the lowest participation ratio, where the participation ratio \cite{wegner_inverse_1980} of a normalized mode $u_n$ is: $PR= 1/(N\sum_n \vert u_n\vert^4)$, indicating their localizated nature. As the gap frequency increases, the topological interface mode persists even when the gap frequencies on the two sides of the interface are not equal, i.e. when $c_L + c_R \neq 2$, as shown in Fig.~\ref{hyperstatic_rotor_chain_sharp_interface}(d).

\emph{Smooth interface.}---To study the modes localized at a smooth interface between the two topologically distinct phases of the hyperstatic rotor chain, we take the continuum limit of Eq.~(\ref{real_space_eom_R_2x1}) with $c_n \to c(x) = 1+m(x)$ and $u_n(t) \to u(x,t)$, $u(x,t)=u(x)\,\mathrm{e}^{i\omega t}$ to get
\begin{align}
& (\omega^2-m^2)u + (1+m)u'' + m' u' = 0  \label{continuum_eom_smooth_interface},
\end{align}
with $u'$ etc. denoting spatial derivatives.

In the limit  $m(x)\ll 1$, i.e., in the region where $c(x)$ is close to $1$, the above equation of motion becomes: $u''(x) + (\omega^2-m^2){u} = 0$, bearing a close resemblance to the time-independent Schr{\"o}dinger equation with the energy $E=\omega^2$, the potential $U(x)=m^2(x) = (c(x)-1)^2$, and $2M/\hbar^2=1$. In this analogy, a smooth interface corresponds to a potential well with a minimum value of $U(x)=0$ when $m(x)=0,\,c(x)=1$, and depth dictated by the asymptotic values of $U(x) = m^2(x)$ on either side of the well. The potential $U(x)$ is symmetric about its minimum when ${m_L}^2 = {m_R}^2$: the asymptotic values of $m^2(x)$ on the left and right side of the interface are equal. This occurs when the lowest bulk mode energies on the two sides of the interface are equal. For a topologically non-trivial interface, this requires that $c_L + c_R = 2$, where $c_L, c_R$ are the asymptotic values of $c(x)$ on the left and right sides of the interface. When this condition is satisfied, at least one bound state solution exists irrespective of the depth of the potential well~\cite{kocher_criteria_1977}, ensuring a localized mode at the interface.

The analogy with the Schr{\"o}dinger equation enables an exact solution of Eq.~(\ref{continuum_eom_smooth_interface}) for $m(x)=m_0 \tanh(x/W)$~\cite{landau_quantum_1981}. For sufficiently large width $W$ and depth $m_0^2$ of the potential well, there are multiple localized modes at each interface, their number given by $\lfloor s(m_0,W)\rfloor +1$, where $s(m_0,W)=(-1+\sqrt{1+(4m_0^2\,W^2)})/2$. The energy of the localized interface modes is $\omega_n^2 = m_0^2-(s(m_0,W)-n)^2/W^2$, for $n=0,1,2,\ldots,\lfloor s\rfloor$, which is less than the lowest bulk mode energy $m_0^2$.

To confirm our exact analysis, we numerically calculate the modes of a periodic hyperstatic chain of $N=4000$ rotors with $m_0=0.3$, $W=6$, and two smooth interfaces for which $m(x)=m_0 \tanh(x/W)$, one of which is shown in Fig.~\ref{hyperstatic_rotor_chain_smooth_interface}(a). The variation of $c(x)=1+m(x)$ (yellow) and the effective potential $U(x)=m^2(x)$ (blue) across the interface is shown in Fig.~\ref{hyperstatic_rotor_chain_smooth_interface}(b). The two localized mode profiles at the interface plotted in Fig.~\ref{hyperstatic_rotor_chain_smooth_interface}(c) (red) are as predicted by the exact solution described in the Appendix. The density of states shown in Fig.~\ref{hyperstatic_rotor_chain_smooth_interface}(d) shows two soft modes (red) per interface having the lowest energies, at the values predicted by the exact solution. These modes also have the lowest participation ratios indicating their localized nature.

The case when the lowest bulk mode energies on the two sides of the interface are unequal, i.e. $c_L + c_R \neq 2$, corresponds to an asymmetric potential well in the analogy with the Schr{\"o}dinger equation. An approximate criterion from Ref.~\cite{kocher_criteria_1977} says that a localized interface mode exists if $W\,(m_L^2 + m_R^2) \gtrsim 2\sqrt{2\vert m_L^2 - m_R^2 \vert}$, where $m_L^2$, $m_R^2$ are the asymptotic values of $m^2(x)$ on the left and right sides of the interface, and $W$ is the interface width. The region of parameter space $\{c_L, c_R \} = \{1+m_L, 1+m_R \}$ for which the above criterion is fulfilled is plotted for different values of $W$ in Fig.~\ref{hyperstatic_rotor_chain_smooth_interface}(e), showing that topological interface modes persist even when $c_L + c_R\neq 2$, with wider and deeper potential wells allowing for greater deviations from the symmetric case.

\textit{Conclusions}---We have presented the first theoretical study of topological mechanical interface modes derived from an overconstrained structure, with exact solutions and numerical calculations. These models rely on generalized inversion symmetry to define topological invariants and create robust soft modes at interfaces between topologically distinct lattices. 
For isostatic topological structures, experimental proposals~\cite{chen_triblock_2011,socolar_mechanical_2017} have yet to be realized on submicron scales. By contrast, the structures that we propose are overconstrained and otherwise rigid, which potentially makes them more accessible to fabrication via existing techniques at scales down to the submicron~\cite{cha_experimental_2018}. Such an architecture will be robust to thermal fluctuations and hence amenable to miniaturization to the micro- and nano-scale. 
Designing topologically protected soft modes in overconstrained materials may lead to future applications from cushioning using soft regions~\cite{bilal2017intrinsically} to controlled failure at topological interfaces~\cite{paulose2015selective}.

\textit{Acknowledgments}---A.S.~acknowledges the support of the Engineering and Physical Sciences Research Council (EPSRC) through New Investigator Award No.~EP/T000961/1. H.K. thanks Jeremy England for his support.

\bibliographystyle{h-physrev}
\bibliography{refs}
\onecolumngrid

\appendix

\section{Generalized Inversion symmetry}

\emph{Generalized inversion symmetry} in a $d-$dimensional mechanical lattice requires (i) that there exist a basis in which the Fourier-transformed rigidity matrix $\mathcal{R}(k)$ is real for real wavenumber $k$ and (ii) that the rigidity matrix in real space be real, as must be the case for a physically realizable  rigidity matrix. Written in terms of the complex wavenumber $z=\mathrm{e}^{ik}$, the rigidity matrix then takes the form $\mathcal{R}(z)=\sum_n{U^\dagger\cdot R_n\,z^n \cdot W}$, where substituting $z=\mathrm{e}^{ik}$ gives the Fourier-transformed rigidity matrix $\mathcal{R}(k) = \mathcal{R}(z)\big\vert_{z=\mathrm{e}^{ik}}$ 

Here, $U, W$ are unitary matrices such that $U\cdot\mathcal{R}(k)\cdot W^\dagger$ is real, hence satisfying condition (i). This requires that $R_n = R_{-n}^\ast$, where $R_n^\ast$ is the complex conjugate of $R_n$. For a zero mode $\mathbf{u}$ at $z=z_0$,
\begin{align}
& \mathcal{R}(z_0)\cdot \mathbf{u} = 0
\quad\Rightarrow \sum_n{ U^\dagger \cdot R_n\,z_0^n\cdot W}\cdot \mathbf{u} = 0  \quad\Rightarrow \sum_n{ R_n\,z_0^n\cdot W}\cdot \mathbf{u} = 0 \quad \Rightarrow \sum_n{ {R_n}^\ast\, (z_0^\ast)^n \cdot W^\ast}\cdot \mathbf{u}^\ast = 0 \nonumber \\
\Rightarrow & \sum_n{ R_{-n}\,(z_0^\ast)^n\cdot W^\ast}\cdot \mathbf{u}^\ast = 0 \quad\Rightarrow \sum_n{ R_{n}\,(z_0^\ast)^{-n}\cdot W^\ast}\cdot \mathbf{u}^\ast = 0  \quad \Rightarrow \sum_n{ R_{n}\,[(z_0^\ast)^{-1}]^n\cdot W^\ast}\cdot \mathbf{u}^\ast = 0  \nonumber \\
\Rightarrow & \sum_n{ \left(U\cdot \mathcal{R}({z_0^\ast}^{-1})\cdot W^\dagger\right) \cdot W^\ast}\cdot \mathbf{u}^\ast = 0  \quad \Rightarrow \sum_n{  \mathcal{R}({z_0^\ast}^{-1})\cdot \big(W^\dagger \cdot W^\ast}\cdot \mathbf{u}^\ast\big) = 0 
\end{align}

Hence, condition (i) implies that for every zero mode $\mathbf{u}$ at $z=z_0$ there is a zero mode $W^\dagger\cdot W^\ast\cdot \mathbf{u}^\ast$ at $z=(z_0^\ast)^{-1}$.

Additionally, condition (ii) requires that $U^\dagger\cdot R_n\cdot W \in \mathbb{R}\,,\, \forall\, n$. For a zero mode $\mathbf{u}$ at $z=z_0$,
\begin{align}
& \mathcal{R}(z_0)\cdot \mathbf{u} = 0
\quad \Rightarrow \sum_n{ U^\dagger \cdot R_n\,z_0^n\cdot W}\cdot \mathbf{u} = 0 \quad\Rightarrow \sum_n{ (z_0^\ast)^n \,\left(U^T \cdot R_n^\ast\cdot W^\ast\right)}\cdot \mathbf{u}^\ast = 0 \nonumber \\
& \Rightarrow \sum_n{ (z_0^\ast)^n\,\left(U^\dagger \cdot R_n\cdot W\right)}\cdot \mathbf{u}^\ast = 0 \quad\Rightarrow  \mathcal{R}(z_0^\ast)\cdot \mathbf{u}^\ast = 0
\end{align}

Hence, condition (ii) implies that for every zero mode $\mathbf{u}$ at $z=z_0$ there is a zero mode $\mathbf{u}^\ast$ at $z=z_0^\ast$. Since generalized inversion symmetry requires both conditions (i) and (ii) be satisfied, a zero mode $\mathbf{u}$ at $z=z_0$ will always be accompanied by a zero mode $\mathbf{u}^\ast$ at $z_0^\ast$, a zero mode $W^\dagger\cdot W^\ast\cdot \mathbf{u}^\ast$ at $z=(z_0^\ast)^{-1}$, and as a consequence of the above, a zero mode  $W^T\cdot W\cdot \mathbf{u}$ at $z=z_0^{-1}$.

Mechanical lattices with generalized inversion symmetry can be classified according to the homotopy groups of their Fourier-transformed rigidity matrix, as in \cite{roychowdhury_classification_2018}. Any matrix, and hence any rigidity matrix $\mathcal{R}(k)$, has a singular value decomposition: $\mathcal{R}=\mathcal{U}\cdot\Lambda_R\cdot\mathcal{V}^\dagger$, and can be continuously transformed into its SVD flattened version $\mathcal{Q}=\mathcal{U}\cdot\tilde{\Lambda}_R\cdot\mathcal{V}^\dagger$ where $\mathcal{U}, \mathcal{V}$ are  unitary matrices, and $\tilde{\Lambda}_R$ is obtained by replacing every element of $\Lambda_R$ by $1$ or $0$ according to whether its magnitude is non-vanishing or vanishing. The SVD-flattened matrix $\mathcal{Q}(k)$ has the same dimensions as $\mathcal{R}(k)$, and encodes the topological properties of $\mathcal{R}(k)$.

In the classification, Maxwell lattices which do not obey generalized inversion symmetry have unitary SVD flattened rigidity matrices, which have a non-trivial fundamental group $\pi_1[U(N)] = \mathbb{Z}$, where the winding number defined by Kane and Lubensky is the integer topological invariant characterizing the different homotopy classes. Maxwell lattices obeying generalized inversion symmetry have orthogonal rigidity matrices with a non-trivial fundamental group $\pi_1[O(N)]$ depending on $N$, which is the dimension of the rigidity matrix in Fourier space, which is the number of sites per unit cell. This classification gives a non-trivial fundamental group for non-Maxwell lattices with `realness' symmetry in which the the number of degrees of freedom and number of constraints per unit cell are mismatched by $\vert \nu\vert = 1$, opening up the possibility of topologically protected modes in non-Maxwell lattices.

In the rest of the Supplemental Material, we show in detail how topologically protected modes arise at interfaces between topologically distinct lattices with generalized inversion symmetry, for the case of one-dimensional Maxwell lattices with $N=2,3$ sites per unit cell, and for a one-dimensional hyperstatic lattice with 2 constraints and 1 degree of freedom per unit cell.

\section{Maxwell Lattices with generalized inversion symmetry}
Maxwell lattices with generalized inversion symmetry have rigidity matrices whose SVD flattened versions that belong to the group of orthogonal matrices $O(N)$. Their fundamental homotopy group: $\pi_1[O(N)]$ is trivial for $N=1$, $\mathbb{Z}$ for $N=2$, and $\mathbb{Z}_2$ for $N\geq 3$, where $N$ is the number of sites per unit cell. 

In this section, we study Maxwell lattices with generalized inversion symmetry belonging to different homotopy classes for $N=2$ and $N=3$, which displays the generic $N>2$ homotopy class, define topological invariants that distinguish these lattices, and show the presence of topologically protected zero modes at interfaces between lattices belonging to different homotopy classes.

\begin{figure}[!htb]
\centering
\includegraphics[width=0.7\columnwidth]{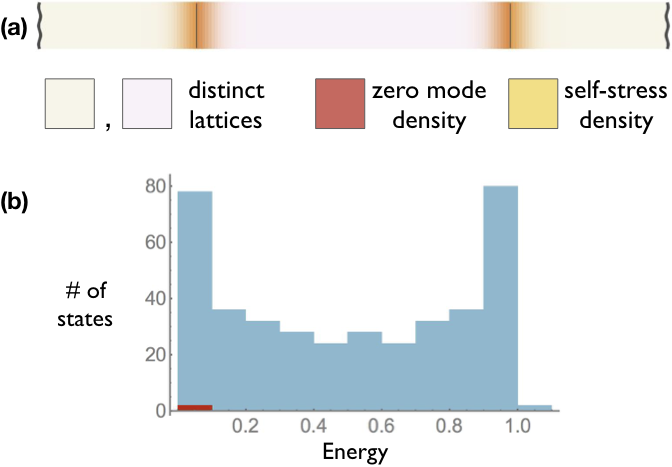}
\caption{Schematic of a material based on a Maxwell lattice with $2$ sites per unit cell, obeying generalized inversion symmetry. The material with the rigidity matrix $\tilde{R}(\lambda, k)$ in Eq.~(\ref{real_space_real_Rmat_2x2}), where $\lambda=0.55$ in the middle region shown in light purple, and $\lambda=0.45$ for the left and right regions shown in light beige. (a) In the presence of generalized inversion symmetry, Maxwell lattices have no topological polarization and consequently have zero modes and self-stress states localized on the both interfaces. (b) The density of states for this system, with the zero modes shown in red.}
\label{2x2_maxwell_gis}
\end{figure}

\subsection{Two sites per unit cell}
In this case, the SVD-flattened rigidity matrix $\mathcal{Q}(k)$ is a $2\times 2$ orthogonal matrix, with determinant $\pm 1$. Multiplying $\mathcal{Q}(k)$ by its determinant gives an $SO(2)$ matrix, which can be classified by the number of times its rotation angle winds around the unit circle, as the wavenumber $k$ goes from $0$ to $2\pi$. The winding number of the rotation angle of the SVD flattened rigidity matrix gives an integer topological invariant distinguishing lattices belonging to different homotopy classes. Since the one-dimensional Brillouin Zone and the rotation angle of a rotation in two dimensions are both defined modulo $2 \pi$, they are both topologically equivalent to a circle, $S^1$, and this homotopy class corresponds to a map of $S^1$ to itself.

An example of such a rigidity matrix is:
\begin{align}
R(\lambda, k) &= -\lambda\,\mathrm{R}(k) + (1-\lambda)\,I =  \begin{pmatrix} (1-\lambda) - \lambda\,\cos(k) & -\lambda\sin(k) \\ +\lambda\sin(k) & (1-\lambda)-\lambda\,\cos(k) \end{pmatrix} \label{2x2_real_rigidity_k}
\end{align}
where $\mathrm{R}(k)$ is a $2\times 2$ rotation matrix with rotation angle $k$. The winding number of the rotation angle of $R(\lambda, k)$ is $0$ or $1$, depending on whether $\lambda$ is less than or greater than $0.5$. 

The above rigidity matrix is real in momentum space, and thus necessarily complex in real space. However, it can be transformed to a rigidity matrix which is real in real space via the following unitary transformation:
\begin{align}
    \tilde{R}(\lambda, k) &= U\cdot R(\lambda,k)\cdot U^\dagger \;,\; U=\frac{1}{2}\begin{pmatrix} 1+i & -1-i \\ 1-i & 1-i \end{pmatrix} \nonumber \\
    \Rightarrow \tilde{R}(\lambda, k) &= \begin{pmatrix} (1-\lambda) - \lambda\,\cos(k) & -i\,\lambda\sin(k) \\ -i\,\lambda\sin(k) & (1-\lambda)-\lambda\,\cos(k) \end{pmatrix} \label{real_space_real_Rmat_2x2}
\end{align}
To further simplify our analysis, the above rigidity matrix is diagonalized via the following orthogonal transformation:
\begin{align}
    R'(\lambda, k) &= O\cdot \tilde{R}(\lambda,k)\cdot O^T \;,\; O=\frac{1}{\sqrt{2}}\begin{pmatrix} 1 & 1 \\ -1 & 1 \end{pmatrix} \nonumber \\
    \Rightarrow R'(\lambda, k) &= \begin{pmatrix} (1-\lambda) - \lambda\,\mathrm{e}^{ik} & 0 \\ 0 & (1-\lambda)-\lambda\,\mathrm{e}^{-ik} \end{pmatrix} \label{diag_Rmat_2x2}
\end{align}
To study the localized zero modes at an interface between lattices belonging to distinct homotopy classes, we rewrite the above rigidity matrix as a function of the complex number $z=\mathrm{e}^{i\,k}$:
\begin{align}
R'(\lambda, z) &= \begin{pmatrix} (1-\lambda) - \lambda\,z & 0 \\ 0 & (1-\lambda)-\lambda z^{-1} \end{pmatrix} \label{2x2_real_rigidity_z}
\end{align}

The determinant of the above rigidity matrix vanishes at $z(\lambda) = \{-\tfrac{\lambda}{1-\lambda}, -\tfrac{1-\lambda}{\lambda} \}$. Notice that the two values of $z$, are of the form $z=z_0, z_0^{-1}$, a pair since $z(\lambda)$ is real, as expected for lattices obeying generalized inversion symmetry. 

This implies that there are left-growing $(\vert z(\lambda)\vert <1)$ and right-growing $(\vert z(\lambda)\vert >1)$ zero modes. The zero modes are orthogonal to each other and are in fact a natural basis for the transformed matrix: $\{(1,0)\,,\, (0,1) \}$. As $\lambda$ crosses $0.5$, the winding number of the rotation angle of the matrix changes from $0 \to 1$, and the zero modes change from $\{$ right growing , left growing $\}$ to $\{$ left growing , right growing $\}$. 

This implies that an interface between two lattices with $\lambda < 0.5$ and $\lambda > 0.5$, would have localized zero modes which would grow at the rates predicted by $z(\lambda)$ on either side of the interface. We have confirmed this observation numerically, by explicitly calculating the zero modes of the dynamical matrix $D=R^T\cdot R$, where $R$ is the real space version of the rigidity matrix in Eq.~(\ref{real_space_real_Rmat_2x2}) for a one-dimensional periodic system with the two topologically distinct phases $\lambda > 0.5, \lambda < 0.5$ spanning $2000$ unit cells each.

Such a real-space matrix may be readily obtained from the previous forms. If a term $c z^n = c \exp(i k n)$ appears in an element of the rigidity matrix it simply corresponds to the same factor $c$ connecting a degree of freedom in any cell indexed $n'$ to one in the cell $n' + n$.

\subsection{Three sites per unit cell}

\begin{figure}[!htb]
\centering
\includegraphics[width=\columnwidth]{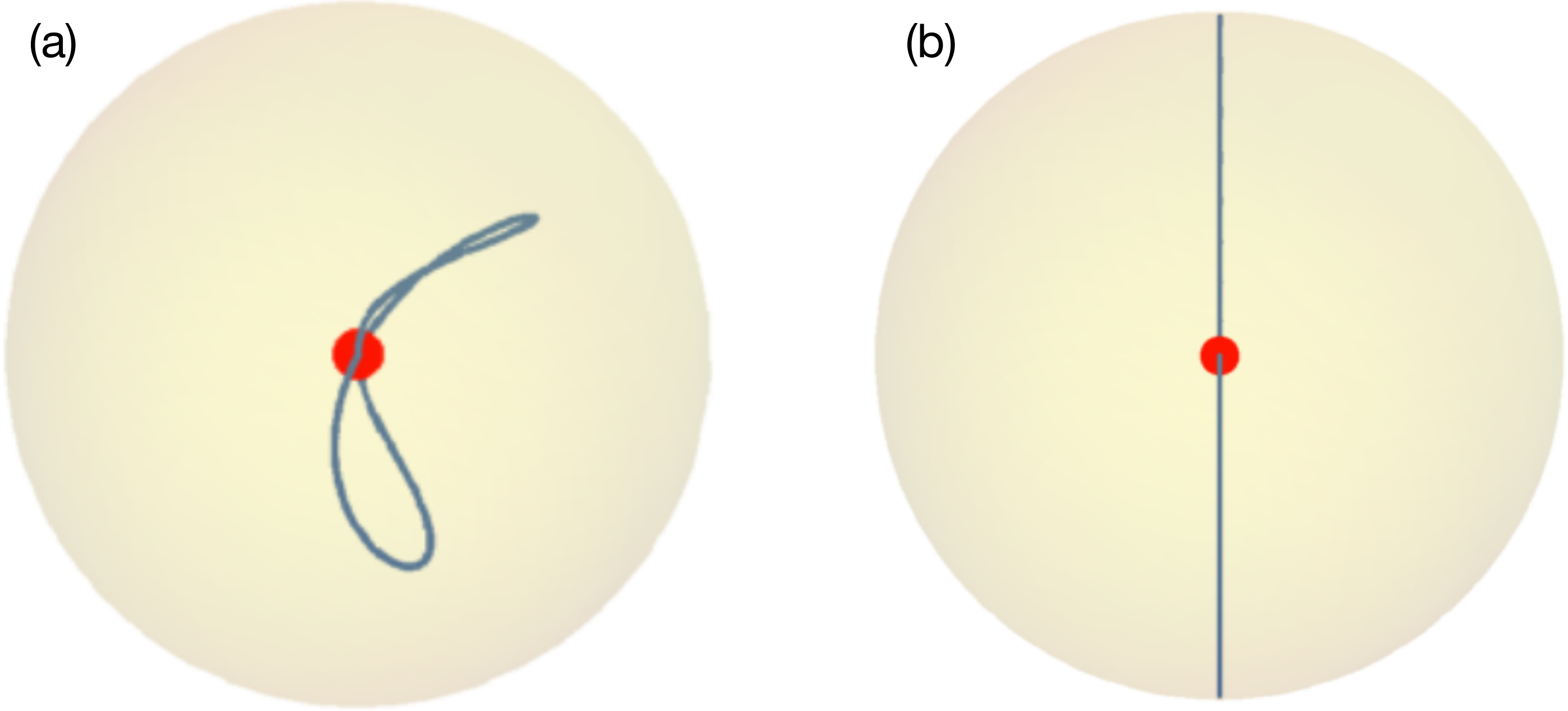}
\caption{(a) a closed loop on the $SO(3)$ sphere which can be continuously shrunk to a point (b) a closed loop on the $SO(3)$ sphere which can not be shrunk to a point, since it closes through antipodal points.}
\label{so3_paths}
\end{figure}

In this case, the SVD-flattened rigidity matrix $\mathcal{Q}(k)$ is a $3\times 3$ orthogonal matrix, with determinant $\pm 1$. Multiplying $\mathcal{Q}(k)$ by its determinant gives an $SO(3)$ matrix for each $k$. Any $SO(3)$ matrix can be represented by a point in a solid sphere (or ``ball'') of radius $\pi$ whose radius vector is along the rotation axis of the matrix, and whose distance from the center of sphere is given by the rotation angle of the matrix~\cite{roychowdhury_classification_2018}. 
Antipodal points on the surface of this solid sphere are identical since a rotation angle of $\pi$ along an axis is the same as a rotation angle of $-\pi$ along the opposite axis. As $k$ varies from $0$ to $2\pi$, the path traced out by the SVD-flattened matrix $\mathcal{Q}(k)\,\in\,SO(3)$ could either be a closed loop which can be shrunk to a point as in Fig.~\ref{so3_paths}(a), or a closed loop which can not be shrunk to a point as in Fig.~\ref{so3_paths}(b) since it closes by piercing the surface of the sphere and coming out through the opposite side. However, a path can be continuously transformed such that pairs of surface crossings cancel out, so that any paths with the same parity of surface crossings are equivalent. Hence the two different homotopy classes of $3\times 3$ rigidity matrices with generalized inversion symmetry are distinguished by a $\mathbb{Z}_2$ topological invariant: the number of times modulo $2$ that the path traced out by the SVD-flattened matrix pierces the surface of the $SO(3)$ sphere where antipodal points are identified.

An example of a topologically non-trivial $3\times 3$ rigidity matrix, inspired by the $2\times 2$ rigidity matrix in the previous section, is: 
\begin{align}
R(\lambda, k) &= -\lambda\,\mathrm{R}_{z}(k) + (1-\lambda)\,I =  \begin{pmatrix} (1-\lambda) - \lambda\,\cos(k) & \lambda\sin(k) & 0 \\ -\lambda\sin(k) & (1-\lambda)-\lambda\,\cos(k) &0 \\
0 & 0 & 1 \end{pmatrix} \label{3x3_real_rigidity_k}
\end{align}
where $\mathrm{R}_{z}(k)$ is a $3\times 3$ rotation matrix with rotation angle $k$ about the z-axis. The number of times that the path traced out by $R(\lambda, k)$ pierces the $SO(3)$ sphere is $0$ or $1$, depending on whether $\lambda$ is less than or greater than $0.5$, as shown in Fig.~\ref{so3_paths}.

Although, we can write the following more general form for a real rigidity matrix:
\begin{align}
R(\lambda, k) &= O(k)\cdot\begin{pmatrix} (1-\lambda) + \lambda\,\cos(k) & -\lambda\sin(k) & 0 \\ \lambda\sin(k) & (1-\lambda)+\lambda\,\cos(k) &0 \\
0 & 0 & 1 \end{pmatrix}\cdot O^T(k)\,;\, O(k)=R_z(\alpha(k))\cdot R_y(\beta(k))\cdot R_x(\gamma(k))  \label{3x3_general_real_rigidity_k}\\
\qquad &  \mathrm{where\;} \; 0\leq \alpha(k),\beta(k),\gamma(k) \leq \pi/2 \,, \quad\mathrm{and}\quad \alpha(0)=\alpha(\pi)=\beta(0)=\beta(\pi)=\gamma(0)=\gamma(\pi)=0   \nonumber 
\end{align}
the path traced out by the above rigidity matrix in the $SO(3)$ sphere can be smoothly transformed to the path traced out by the rigidity matrix given in Eq.~(\ref{3x3_real_rigidity_k}), and hence will have the same topological invariant as the rigidity matrix in Eq.~(\ref{3x3_real_rigidity_k}). Since the topology of the above more general rigidity matrix is captured by the simpler rigidity matrix given in Eq.~(\ref{3x3_real_rigidity_k}), we proceed to study the simpler case.

The rigidity matrix though complex in real space, is easily transformed to a rigidity matrix which is real in real space via the following unitary transformation: 
\begin{align}
    \tilde{R}(\lambda, k) &= U\cdot R(\lambda,k)\cdot U^\dagger \;,\; U=\begin{pmatrix} (1+i)/2 & (-1-i)/2 & 0 \\ (1-i)/2 & (1-i)/2 & 0 \\ 0 & 0 & 1  \end{pmatrix} \nonumber \\
    \Rightarrow \tilde{R}(\lambda, k) &= \begin{pmatrix} (1-\lambda) + \lambda\,\cos(k) & i\,\lambda\sin(k) & 0 \\ i\,\lambda\sin(k) & (1-\lambda)+\lambda\,\cos(k) & 0 \\
    0 & 0 & 1\end{pmatrix} \label{real_space_real_Rmat_3x3}
\end{align}
To further simplify our analysis, we diagonalize the above rigidity matrix via the following orthogonal transformation:
\begin{align}
    R'(\lambda, k) &= O\cdot \tilde{R}(\lambda,k)\cdot O^T \;,\; O=\begin{pmatrix} 1/\sqrt{2} & 1/\sqrt{2} & 0 \\ -1/\sqrt{2} & 1/\sqrt{2} & 0 \\ 0 & 0 & 1 \end{pmatrix} \nonumber \\
    \Rightarrow R'(\lambda, k) &= \begin{pmatrix} (1-\lambda) + \lambda\,\mathrm{e}^{ik} & 0 & 0 \\ 0 & (1-\lambda)+\lambda\,\mathrm{e}^{-ik} & 0 \\ 0 & 0 & 1 \end{pmatrix} \label{diag_Rmat_3x3}
\end{align}
To study the localized zero modes at an interface between lattices belonging to distinct homotopy classes, we rewrite the above rigidity matrix as a function of the complex number $z=\mathrm{e}^{i\,k}$:
\begin{align}
R'(\lambda, z) &= \begin{pmatrix} (1-\lambda) + \lambda\,z & 0 & 0 \\ 0 & (1-\lambda)+\lambda z^{-1} & 0 \\ 0 & 0 & 1 \end{pmatrix} \label{3x3_real_rigidity_z}
\end{align}

The determinant of the above rigidity matrix vanishes at $z(\lambda) = \{ -\tfrac{\lambda}{1-\lambda}, -\tfrac{1-\lambda}{\lambda} \}$, where the product of the two roots is 1, at values of $z$ of the form $z=z_0, z_0^{-1}$, a pair since $z(\lambda)$ is real, as expected for lattices with generalized inversion symmetry. This implies that there is a left growing $(\vert z(\lambda)\vert <1)$ and a right growing $(\vert z(\lambda)\vert >1)$ zero mode. The zero modes are orthogonal to each other: $\{(1,0,0)\,,\, (0,1,0) \}$. As $\lambda$ crosses $0.5$, the number of times the path traced out by the rigidity matrix pierces the $SO(3)$ sphere changes from $0 \to 1$, and the zero modes change from $\{$ left growing $\,,\,$ right growing $\}$ to $\{$ right growing $\,,\,$ left growing $\}$.

This implies that an interface between lattices with $\lambda < 0.5$ and $\lambda > 0.5$ would have localized zero modes which would grow at the rates predicted by $z(\lambda)$ on either side of the interface. This is confirmed numerically.

\section{A hyperstatic lattice with generalized inversion symmetry}

For a hyperstatic lattice with generalized inversion symmetry, with 2 springs and 1 degree of freedom per unit cell, there is a choice of basis in which the Fourier transform of the rigidity matrix will be a real $2\times 1$ matrix, which can be thought of as a real two-vector.
The space of $2\times 1$ rigidity matrices can be classified according to the fundamental homotopy group of their SVD flattened versions, giving $\mathbb{Z}$ distinct homotopy classes characterized by a integer topological invariant: the winding number of the two-vector $\vec{R}(k)$ as $k$ goes from $0\to 2\pi$.

An example of such a Fourier transformed rigidity matrix is:
\begin{equation}
    R(k) = \begin{pmatrix} c - \cos(k) \\ \sin(k) \end{pmatrix} \label{bulk_real_rigidity_2x1_fourier}
\end{equation}
The winding number of the above matrix goes from $1$ to $0$ as $\vert c \vert$ goes from $\vert c\vert <1$ to $\vert c \vert >1$. As $\vert c\vert$ passes through $1$, the bulk energy gap closes, i.e. the lowest energy: $\min_k \{R^\dagger(k) R(k)\}=\min_k\{1+c^2-2c\cos(k)\}=(1-\vert c\vert)^2$ goes to zero at $\vert c\vert=1$, and is positive otherwise.

Since the above rigidity matrix is real in momentum space, it is necessarily complex in real space. However, it can be easily transformed to a rigidity matrix that is real in real space via the following unitary transformation:
\begin{align}
    \mathcal{R}(k) &= U\cdot R(k) = \begin{pmatrix} 1 & 0 \\ 
    0 & i \end{pmatrix} \cdot \begin{pmatrix} c - \cos(k) \\ \sin(k) \end{pmatrix} \nonumber \\
\Rightarrow \mathcal{R}(k) &= \begin{pmatrix} c-\cos(k) \\  i\,\sin(k) \end{pmatrix} \label{real_space_real_rigidity_2x1_fourier}
\end{align}

The above matrix can be further simplified via an orthogonal transformation:
\begin{align}
    \tilde{R}(k) &= O\cdot \mathcal{R}(k) = \begin{pmatrix} \frac{1}{\sqrt{2}} & \frac{-1}{\sqrt{2}} \\ 
    \frac{1}{\sqrt{2}} & \frac{1}{\sqrt{2}}
    \end{pmatrix} \cdot \begin{pmatrix} c - \cos(k) \\ i\,\sin(k) \end{pmatrix} \nonumber \\
   \Rightarrow \tilde{R}(k) &= \frac{1}{\sqrt{2}}\begin{pmatrix} c-\mathrm{e}^{i\,k} \\ 
    c-\mathrm{e}^{-i\,k}
    \end{pmatrix} \label{rotated_real_rigidity_2x1_fourier}
\end{align}
which can be realized by a chain of rotors connected by springs shown in Fig.~2 of the manuscript, for which the rigidity matrix $\mathcal{R}'(k)$ is:
\begin{align}
\mathcal{R}'(k) &= \frac{a-2r\sin\theta}{\sqrt{a^2+4r^2\cos^2\theta}} \begin{pmatrix} \frac{a+2r\sin\theta}{a-2r\sin\theta} - \mathrm{e}^{ik} \\
\frac{a+2r\sin\theta}{a-2r\sin\theta} - \mathrm{e}^{-ik} \end{pmatrix}  \label{hyperstatic_rotor_chain_Rmat}
\end{align}
where $a$ is the lattice spacing, $r$ is the distance between the fixed point and the rotor head, and $\theta$ is the rotor angle measured from the vertical for the rotor head in red, as shown in Fig.~2(a),(c) of the manuscript. $\tilde{R}(k)\equiv\mathcal{R}'(k)$ up to the multiplication of a scalar, with $c=(a+2r\sin\theta)/(a-2r\sin\theta)$.

We now derive the dynamical equation of motion in real space for the rigidity matrix given in Eq.~(\ref{rotated_real_rigidity_2x1_fourier})

The rigidity matrix in real space corresponding to the Fourier transformed matrix given in Eq.~\ref{bulk_real_rigidity_2x1_fourier} is given by:
\begin{align}
& \begin{pmatrix} R_{2j-1,j'} \\ R_{2j, j'} \end{pmatrix} = \frac{1}{\sqrt{2}}\begin{pmatrix} c_j\delta_{j,j'} -\delta_{j,j'+1} \\ c_j\delta_{j,j'} - \delta_{j,j'-1} \end{pmatrix} \label{real_space_rotated_R_2x1}
\end{align}

The corresponding real space equation of motion is given by:
\begin{align}
\ddot{\mathbf{u}} &= - \overset{\text{\tiny$\leftrightarrow$}}{\mathbf{D}}\cdot \mathbf{u} = - \mathbf{R}^T\cdot\mathbf{R}\cdot\mathbf{u} \nonumber \\
\Rightarrow \ddot{u}_n &= -\sum_{j,l} R^T_{n,j}\,R_{j,l}\,u_l = - \sum_{j,l} R_{j,n}\,R_{j,l}\,u_l \nonumber \\
&= -\sum_{j=1}^N \sum_{l=1}^N \left( R_{2j-1,n} R_{2j-1,l} + R_{2j,n} R_{2j,l} \right) u_l \nonumber 
\\
&= -\frac{1}{2}\sum_{j=1}^N \sum_{l=1}^N \left( \left(c_j \delta_{j,n} - \delta_{j,n+1} \right) \left( c_j \delta_{j,l} - \delta_{j,l+1} \right) + \left(c_j\delta_{j,n} -\delta_{j,n-1} \right)\left( c_j \delta_{j,l} - \delta_{j,l-1} \right) \right) u_l \nonumber \\
&= -\sum_{j=1}^N\sum_{l=1}^N \left( c_j^2 \delta_{j,n}\delta_{j,l} + \frac{1}{2}(\delta_{j,n+1}\delta_{j,l+1}+\delta_{j,n-1}\delta_{j,l-1}) - \frac{c_j}{2}(\delta_{j,l}\delta_{j,n+1} + \delta_{j,l}\delta_{j,n-1} + \delta_{j,n}\delta_{j,l+1} + \delta_{j,n}\delta_{j,l-1}) \right)u_l \nonumber \\
&= -\sum_{l=1}^N \left( (c_n^2+1)\delta_{n,l} -\frac{1}{2}\sum_{j=1}^N c_j(\delta_{j,l}\delta_{j,n+1} + \delta_{j,l}\delta_{j,n-1}+\delta_{j,n}\delta_{j,l+1}+\delta_{j,n}\delta_{j,l-1}) \right)u_l\nonumber \\
&= -\sum_{l=1}^N\left( (c_n^2+1)\delta_{n,l} - \frac{1}{2}(c_l\delta_{l,n+1}+c_l\delta_{l,n-1}+c_n\delta_{n,l+1} + c_n\delta_{n,l-1})\right) u_l\nonumber \\
&=-\left( (c_n^2+1)u_n - \frac{1}{2}(c_{n+1}u_{n+1}+c_{n-1}u_{n-1}+c_n u_{n-1} + c_n u_{n+1}) \right) \nonumber\\
&= -(c_n-1)^2u_n +\left( \frac{c_{n+1}+c_n}{2}u_{n+1} - 2c_n u_n +\frac{c_n + c_{n-1}}{2}u_{n-1}  \right) \nonumber \\
\Rightarrow \ddot{u}_n &= -(c_n-1)^2u_n +c_n ( u_{n+1} - 2 u_n +u_{n-1} ) +\frac{c_{n+1}-c_n}{2} u_{n+1} -\frac{c_n - c_{n-1}}{2} u_{n-1} \label{real_space_eom_R_2x1}
\end{align}

Using the above equation of motion, we will show that at an interface between topologically distinct lattices, i.e. where $c$ crosses $1$, irrespective of whether the interface is smooth or sharp, a soft mode localized at the interface always exists when the values of $c$ on either side of the interface are symmetric about 1, i.e. when the values of $c$ on either side sum to 2. For the general case of an interface where $c$ crosses $1$, but is not symmetric about $1$ on either side, we will solve for the conditions for the existence of localized soft modes, for both smooth and sharp interfaces.

We will treat the case of the smooth and sharp interface separately, beginning with a sharp interface.

\subsection{Sharp Interface}

We consider a sharp interface between a lattice with uniform $c=c_L$ on the left and $c=c_R$ on the right. For the bulk regions on either side, writing $u_n(t) = u_0\, z^n\,\mathrm{e}^{i\omega t}$, where $u_0$ is the displacement at the site closest to the interface, the equation of motion: Eq.~(\ref{real_space_eom_R_2x1}) becomes:
\begin{align}
-\omega^2\,u_0\,z^n &= -(c -1)^2 u_0\,z^n + c\,u_0\,z^n(z-2+z^{-1}) \nonumber \\
\Rightarrow \omega^2 &= (c-1)^2 - c(z+z^{-1}-2) \nonumber \\
\Rightarrow \omega^2 &= (c_L-1)^2 - c_L(z_L +z^{-1}_L-2) \label{left_bulk} \\
\Rightarrow \omega^2 &= (c_R-1)^2 - c_R(z_R +z^{-1}_R-2) \label{right_bulk}
\end{align}
where $z_L,z_R$ are complex numbers describing the growth, decay or oscillatory nature of the mode to the left and right of the interface depending on whether $\vert z_{L(R)}\vert$ is greater than, less than, or equal to $1$.

Labeling the displacements at the sites adjacent to the interface as $u_{0,L}$ on the left, and $u_{0,R}$ on right, and rewriting Eq.~(\ref{real_space_eom_R_2x1}) for these sites, we get:
\begin{align}
\omega^2 u_{0,L} &= (c_L-1)^2 u_{0,L} - c_L\left( (z_L^{-1}-2)u_{0,L} + u_{0,R} \right) - \frac{c_R - c_L}{2}u_{0,R} \label{left_interface_site} \\
\omega^2 u_{0,R} &= (c_R-1)^2 u_{0,R} - c_R\left( (z_R-2)u_{0,R} + u_{0,L} \right) + \frac{c_R - c_L}{2}u_{0,L} \label{right_interface_site}
\end{align}
Substituting for $\omega^2$ from Eq.s~(\ref{left_bulk}),(\ref{right_bulk}) into Eq.s~(\ref{left_interface_site}),(\ref{right_interface_site}), we get:
\begin{align}
 c_L z_L u_{0,L} - c_L u_{0,R} - \frac{c_R-c_L}{2}u_{0,R} &= 0 \label{interface_eq_1} \\
 c_R z_R^{-1} u_{0,R} - c_R u_{0,L} + \frac{c_R - c_L}{2}u_{0,L} &= 0 \label{interface_eq_2}
\end{align}
For a non-trivial solution to exist for Eq.s~(\ref{interface_eq_1}),(\ref{interface_eq_2}), the following determinant must vanish:
\begin{align}
\det\begin{pmatrix} c_L z_L & -\frac{c_R+c_L}{2} \\ -\frac{c_R+c_L}{2} & c_R z_R^{-1} \end{pmatrix} &= 0 \nonumber \\
\Rightarrow \frac{z_L}{z_R} =\frac{(c_R+c_L)^2}{4\,c_R\, c_L} \label{interface_mode_decay_ratio}
\end{align}
Notice from Eq.~(\ref{interface_mode_decay_ratio}) that $z_L/z_R-1>0\Rightarrow z_L/z_R>1$.

We now have three equations: Eq.s~(\ref{left_bulk}),(\ref{right_bulk}),(\ref{interface_mode_decay_ratio}) for three variables: $z_R, z_L, \omega^2$. 

Subtracting Eq.~(\ref{left_bulk}) from Eq.~(\ref{right_bulk}), eliminates $\omega^2$ and gives:
\begin{align}
(c_R-1)^2 - (c_L-1)^2 - c_R(z_R+z_R^{-1}-2) + c_L(z_L + z_L^{-1}-2) &= 0 \nonumber \\
\Rightarrow (c_R^2 - c_L^2) - c_R(z_R+z_R^{-1}) + c_L(z_L+z_L^{-1}) &= 0 \label{interface_mode_decay_reln}
\end{align}

Using Eq.~(\ref{interface_mode_decay_ratio}) to eliminate $z_L$ in favor of $z_R$, we can solve for $z_R$ as follows:
\begin{align}
(c_R^2 - c_L^2) - c_R(z_R + z_R^{-1}) + c_L\left( \frac{(c_L+c_R)^2}{4c_L c_R} z_R + \frac{4c_L c_R}{(c_L + c_R)^2} z_R^{-1} \right) &= 0 \nonumber \\
\Rightarrow \left(1-\frac{(c_L+c_R)^2}{4c_R^2}\right)z_R^2 - \frac{c_R^2 - c_L^2}{c_R}z_R + \left( 1-\frac{4c_L^2}{(c_L+c_R)^2}\right) &=0 \nonumber \\
\Rightarrow z_R = \frac{\frac{c_R^2-c_L^2}{c_R}\pm \sqrt{\left(\frac{c_R^2-c_L^2}{c_R}\right)^2 -4\left(1-\frac{(c_L+c_R)^2}{4c_R^2} \right)\left(1-\frac{4 c_L^2}{(c_L+c_R)^2}\right)}}{2\left( 1-\frac{(c_L+c_R)^2}{4c_R^2}\right)} \label{zR_roots}
\end{align}
Substituting the values for $z_R$ from Eq.~(\ref{zR_roots}) into Eq.~(\ref{interface_mode_decay_ratio}) solves for $z_L$, and substitution into Eq.~(\ref{right_bulk}) solves for $\omega^2$.

For the solution to be a localized interface mode, it must be growing on the left side: $\vert z_L \vert >1$ and decaying on the right side: $\vert z_R \vert < 1$. In the region of parameter space $0<c_L,c_R<2$, the values of $\{c_L,c_R \}$ which give localized interface modes are plotted in Fig.~3(d) of the main text. 

For the case when $c_L,c_R$ are symmetric about $1$, taking $c_L=1-m_0,\,,c_R=1+m_0$, we get the following values for $z_R,z_L,\omega^2$:
\begin{align}
z_R &= \frac{(1+m_0)(2-m_0)}{2+m_0}\,,\, z_L = \frac{2-m_0}{(2+m_0)(1-m_0)}\,,\,\omega^2=m_0^2 - \frac{m_0^4}{4-m_0^2} \label{sharp_interface_symmetric_soln}
\end{align}
The above values of the mode growth/decay rates and the mode energy are confirmed numerically.

\subsection{Smooth Interface}
For a smooth interface, we take the continuum limit of Eq.~(\ref{real_space_eom_R_2x1}) with $u_n(t) \to u(x,t)$. Assuming $u(x,t)=u(x)\,\mathrm{e}^{i\omega t}$, $\ddot{u}(x,t)=-\omega^2\,u(x)\,\mathrm{e}^{i\omega t}$, Eq.~(\ref{real_space_eom_R_2x1}) becomes:
\begin{align}
-\omega^2{u}(x) &= -(c(x)-1)^2 u(x) + c(x)u''(x) + c'(x) u'(x)  \label{continuum_eom_smooth_interface}
\end{align}
Taking $c(x)=1+m(x)$, and working in the limit where $m(x)\ll 1$, the above continuum equation of motion becomes:
\begin{align}
&-\omega^2{u}(x) \approx -m^2(x) u(x) + u''(x) \nonumber \\
&\Rightarrow  u'' + (\omega^2 - m^2(x))u = 0 \label{approx_continum_smooth}
\end{align}
which can be mapped to the time-independent Schr{\"o}dinger equation:
\begin{equation}
\psi'' + \frac{2m}{\hbar^2}(E-U(x))\psi = 0 \label{schrodinger}
\end{equation}
with $\frac{2m}{\hbar^2}=1$, and $E-U(x)=\omega^2-m^2(x)$.

At a smooth interface where $c(x)$ crosses $1$, the potential energy given by $m^2(x)$ drops down smoothly to $0$ at the interface and climbing up to the bulk values of $(c-1)^2$ on either side of the interface. Hence, a smooth interface between topologically distinct lattices implies a potential well at the interface in the above mapping between Eq.~(\ref{approx_continum_smooth}) and Eq.~(\ref{schrodinger}), since $c(x)$ must cross $1$ at such an interface. When the asymptotic values of the potential well on the two sides are equal, at least one bound state solution for the Schr{\"o}dinger equation, Eq.~(\ref{schrodinger}), exists irrespective of the depth of the potential well as shown in \cite{kocher_criteria_1977}. However, when the potential well is asymmetric, i.e. the asymptotic values of the potential energy on either side of the well are unequal, an approximate criterion for a bound state to exist is given by \cite{kocher_criteria_1977} as:
\begin{equation}
W\,V_0 \gtrsim \sqrt{2\Delta V} \label{smooth_interface_bound_state_criterion}
\end{equation}
where $W$ is the width of the potential well, $V_0$ is the depth of the potential well with respect to average of the potential energy on either side of the well, and $\Delta V$ is the difference between the potential energies on either side of the well.

Using the above criterion, a localized soft mode at a smooth interface where $c$ crosses $1$ exists when 
\begin{equation}
W\,\frac{m_L^2 + m_R^2}{2} \gtrsim \sqrt{2\vert m_L^2 - m_R^2 \vert} \label{smooth_interface_localized_mode_criterion}
\end{equation}
where $m_L^2, m_R^2$ are the bulk values of $m^2(x)=(c(x)-1)^2$ on the left and right sides of the interface, and $W$ is the width of the region over which $c(x)$ varies. The region of parameter space specified by Eq.~(\ref{smooth_interface_localized_mode_criterion}) is plotted in Fig.~4(e) of the manuscript. 

An explicit example of a mapping between Eq.~(\ref{approx_continum_smooth}) and Eq.~(\ref{schrodinger}) is possible for $m(x)=m_0 \tanh(x/W)$, where the width of the potential well $m^2(x) = m_0^2\left(1-1/{\cosh^2(x/W)}\right)$ is given by $W$. Since $m^2(x) = m_0^2\left(1-1/{\cosh^2(x/W)}\right)$, the equation of motion becomes:
\begin{equation}
u''(x) + \left((\omega^2-m_0^2) + \frac{m_0^2}{\cosh^2(x/W)}\right)u(x) = 0 \label{sech_squared_m(x)}
\end{equation}
which can be mapped to the Schr{\"o}dinger equation with $E=(\omega^2-m_0^2)$, and the potential $U(x)=-m_0^2/\cosh^2(x/W)$.

The number of localized modes, i.e. with $E<0$, or equivalently with $\omega^2 < m_0^2$ is given by $\lfloor s(m_0,W)\rfloor +1$, where $s(m_0,W)=\frac{-1+\sqrt{1+(4m_0^2\,W^2)}}{2}$. 

The eigenvalues of the dynamical matrix for the localized modes are given by $\omega_n^2 = m_0^2-(s(m_0,W)-n)^2/W^2$, for $n=0,1,2,\ldots,\lfloor s\rfloor$.

The localized eigenmodes are given by:
\begin{align}
u_n(x) &= \left(1-\tanh^2\left(\frac{x}{W}\right)\right)^\frac{s(m_0,W)-n}{2}\,\pFq{2}{1}\left[-n,\, 2s(m_0,W)+1-n;\,s(m_0,W)-n+1;\,\tfrac{1}{2}\left(1-\tanh\frac{x}{W}\right)\right]  \label{exact_localized_modes_smooth}
\end{align}

Note that $m_0\to 0\,, W \sim 1\Rightarrow s \to 0^+$, and the analytical form of the localized eigenmodes approaches $\pFq{2}{1}[0,\,1;\,1;\,\frac{1}{2}(1-\tanh(x/W))]$ which is a flat line, i.e. a fully delocalized mode.

Examining the regions of parameter space where this approximation breaks:
\begin{enumerate}
    \item Large $m_0$ $(m_0\gtrsim  1)$.

The approximation required to arrive at Eq.~(\ref{approx_continum_smooth}) is no longer valid, since the neglected term $m\,u''$ is now comparable to the retained term $u''$. The number of predicted localized modes and the shape of the predicted localized modes are no longer in agreement with the numerically calculated modes, however the numerically calculated lowest energy modes are localized at the interface. 

\item Narrow interface ($W\sim 1$).

Even though interface between the topologically distinct bulk phases is now much sharper, where the continuum approximation is not expected to hold, the numerically calculated localized modes match the analytically predicted modes. As $m_0\rightarrow 0$, both the numerically calculated modes and the analytically predicted modes become delocalized. 
\end{enumerate}

\end{document}